\documentstyle[epsfig,12pt,a4p]{article}

\newcommand{\pipi}{\mbox{$\pi^+\pi^-$ }}

\begin{document}
\begin{titlepage}
\def\footnoterule{\hrule width 1.0\columnwidth}
\begin{tabbing}
put this on the right hand corner using tabbing so it looks
 and neat and in \= \kill
\> {8 September 1999}
\end{tabbing}
\bigskip
\bigskip
\begin{center}{\Large  {\bf Experimental evidence for a vector-like
behaviour of Pomeron exchange}}
\end{center}
\bigskip
\bigskip
\begin{center}{        The WA102 Collaboration
}\end{center}\bigskip
\begin{center}{
D.\thinspace Barberis$^{  4}$,
F.G.\thinspace Binon$^{   6}$,
F.E.\thinspace Close$^{  3,4}$,
K.M.\thinspace Danielsen$^{ 11}$,
S.V.\thinspace Donskov$^{  5}$,
B.C.\thinspace Earl$^{  3}$,
D.\thinspace Evans$^{  3}$,
B.R.\thinspace French$^{  4}$,
T.\thinspace Hino$^{ 12}$,
S.\thinspace Inaba$^{   8}$,
A.\thinspace Jacholkowski$^{   4}$,
T.\thinspace Jacobsen$^{  11}$,
G.V.\thinspace Khaustov$^{  5}$,
J.B.\thinspace Kinson$^{   3}$,
A.\thinspace Kirk$^{   3}$,
A.A.\thinspace Kondashov$^{  5}$,
A.A.\thinspace Lednev$^{  5}$,
V.\thinspace Lenti$^{  4}$,
I.\thinspace Minashvili$^{   7}$,
J.P.\thinspace Peigneux$^{  1}$,
V.\thinspace Romanovsky$^{   7}$,
N.\thinspace Russakovich$^{   7}$,
A.\thinspace Semenov$^{   7}$,
P.M.\thinspace Shagin$^{  5}$,
H.\thinspace Shimizu$^{ 10}$,
A.V.\thinspace Singovsky$^{ 1,5}$,
A.\thinspace Sobol$^{   5}$,
M.\thinspace Stassinaki$^{   2}$,
J.P.\thinspace Stroot$^{  6}$,
K.\thinspace Takamatsu$^{ 9}$,
T.\thinspace Tsuru$^{   8}$,
O.\thinspace Villalobos Baillie$^{   3}$,
M.F.\thinspace Votruba$^{   3}$,
Y.\thinspace Yasu$^{   8}$.
}\end{center}

\begin{center}{\bf {{\bf Abstract}}}\end{center}

{
Evidence is presented that the Pomeron act as a non-conserved vector current.
A study has been made of
the azimuthal angle $\phi$, which is defined as the angle between the $p_T$
vectors of the two outgoing protons,
in the reaction $pp \rightarrow pp(X^0)$
for those resonances ($X^0$)
which are compatible with
being produced by double Pomeron exchange.
These distributions
have been compared
with a model which describes the
Pomeron as a non-conserved vector current and a
qualitative agreement is found.
In addition,
when one of the particles exchanged is known to have spin 0, namely
$\pi$-Pomeron exchange, the $\phi$ distribution is flat.
}
\bigskip
\bigskip
\bigskip
\bigskip\begin{center}{{Submitted to Physics Letters}}
\end{center}
\bigskip
\bigskip
\begin{tabbing}
aba \=   \kill
$^1$ \> \small
LAPP-IN2P3, Annecy, France. \\
$^2$ \> \small
Athens University, Physics Department, Athens, Greece. \\
$^3$ \> \small
School of Physics and Astronomy, University of Birmingham, Birmingham, U.K. \\
$^4$ \> \small
CERN - European Organization for Nuclear Research, Geneva, Switzerland. \\
$^5$ \> \small
IHEP, Protvino, Russia. \\
$^6$ \> \small
IISN, Belgium. \\
$^7$ \> \small
JINR, Dubna, Russia. \\
$^8$ \> \small
High Energy Accelerator Research Organization (KEK), Tsukuba, Ibaraki 305-0801,
Japan. \\
$^{9}$ \> \small
Faculty of Engineering, Miyazaki University, Miyazaki 889-2192, Japan. \\
$^{10}$ \> \small
RCNP, Osaka University, Ibaraki, Osaka 567-0047, Japan. \\
$^{11}$ \> \small
Oslo University, Oslo, Norway. \\
$^{12}$ \> \small
Faculty of Science, Tohoku University, Aoba-ku, Sendai 980-8577, Japan. \\
\end{tabbing}
\end{titlepage}
\setcounter{page}{2}
\bigskip
\par
Experiment WA102 is designed to study exclusive final states formed in
the reaction
\begin{equation}
pp \rightarrow p_{f} (X^0) p_{s}
\end{equation}
at 450 GeV/c.
The subscripts $f$ and $s$ indicate the
fastest and slowest particles in the laboratory respectively
and $X^0$ represents the central
system that is presumed to be produced by double exchange processes:
in particular Double Pomeron Exchange (DPE).
The experiment
has been performed using the CERN Omega Spectrometer,
the layout of which is
described in ref.~\cite{WADPT}.
In previous analyses it has been observed that
when the centrally produced system has been analysed
as a function of the parameter $dP_T$, which is the difference
in the transverse momentum vectors of the two exchange
particles~\cite{WADPT,closeak},
all the undisputed
$ q \overline q $ states
(i.e. $\eta$, $\eta^{\prime}$, $f_1(1285)$ etc.)
are suppressed as $dP_T$ goes to zero,
whereas the glueball candidates
$f_0(1500)$, $f_0(1710)$ and $f_2(1950)$ are prominent~\cite{memoriam}.
\par
In addition, an interesting effect has been observed in
the azimuthal angle $\phi$ which is defined as the angle between the $p_T$
vectors of the two outgoing protons.
Historically it has been assumed that the Pomeron, with ``vacuum quantum
numbers", transforms as a scalar and hence
that the $\phi$ distribution
would be flat for resonances
produced by DPE.
The $\phi$ dependences observed~\cite{ourpubs,0mppap,f1pap,pipikk,pipipi0}
are clearly not flat and considerable variation
is observed among the resonances produced.
\par
Several theoretical papers have been published on these
effects~\cite{angdist,clschul}.
All agree that the exchanged particle
must have J~$>$~0
and that J~=~1
is the simplest explanation for the observed $\phi$ distributions.
Close and Schuler~\cite{clschul} have calculated the $\phi$ dependences
for the production of resonances with different $J^{PC}$
for the case where the exchanged particle is a
Pomeron that transforms like a non-conserved
vector current.
In order
to try to get some insight into the nature of the particles
exchanged in central $pp$ interactions
we will compare
the predictions of this model with the data
for resonances with different $J^{PC}$ observed in the
WA102 experiment.
\par
The simplest situation is for the production of
$J^{PC}$~=~$0^{-+}$ states where the model of Close and Schuler~\cite{clschul}
predicts
\begin{equation}
\frac{d^3\sigma}{d\phi dt_1dt_2} \propto t_1 t_2 \sin^2 \phi
\label{eq:a}
\end{equation}
where $t_1$ and $t_2$ are the
four momentum transfer at the beam-fast and target-slow vertices
respectively.
Fig.~\ref{fi:1}a) and b)
show the
experimental $\phi$ distributions for the $\eta$ and $\eta^\prime$.
They have been fitted to the form $\alpha \sin^2 \phi $ which
describes the data well.
It has also been found
experimentally
that $d\sigma/dt$ is proportional
to $t$ ~\cite{0mppap} (where $t$ is $t_1$ or $t_2$)
as predicted from equation~(\ref{eq:a}).
\par
The fact that
both the $\eta$ and $\eta^{\prime}$ signals are suppressed at small
four-momentum transfers, where Double Pomeron Exchange (DPE) is believed
to be dominant, was assumed to imply
that the
$0^{-+}$ states do not couple to DPE~\cite{eiota}.
However, from equation~(\ref{eq:a})
it can be seen that if DPE is mediated via Pomerons transforming as
vector particles then the production of
$0^{-+}$ resonances will be suppressed at small $t$.
Equation~(\ref{eq:a}) is general to all vector vector exchange
processes, so to investigate if the $\eta$ and $\eta^{\prime}$ are
produced by DPE we have attempted to determine their cross sections
as a function of energy. To do so we have calculated
the ratio of the cross
sections
measured by the WA76 experiment,
at 85~GeV/c ($\sqrt{s}$~=~12.7~GeV), to those
measured by the WA102 experiment
at 450~GeV/c ($\sqrt{s}$~=~29.1~GeV).
\par
Up to now
the determination of this ratio has been limited to
resonances that decay to final states containing only
charged particles
due to
the fact that there was no calorimeter
in the 85~GeV/c run of the WA76 experiment. However,
the WA76 experiment was able to reconstruct the
$\eta \pi^+\pi^-$ mass spectrum using the
decay $\eta$~$\rightarrow$~$\pi^+\pi^-(\pi^0)_{missing}$~\cite{WA76etapipi}.
In this mass spectrum the $\eta^\prime$ and $f_1(1285)$ are seen.
The cross section of the $f_1(1285)$ at 85 GeV/c has been well
measured through its all charged particle decay mode
and hence can be used to determine the
cross section of the $\eta^\prime$ after taking into account the
different acceptance and combinatorial effects and gives
\begin{equation}
\frac{\sigma_{450}(\eta^\prime)}{\sigma_{85}(\eta^\prime)} = 0.72 \pm 0.16
\end{equation}
For Pomeron-Pomeron exchange we would expect a value of $\approx$ 1.0,
while for $\rho$-$\rho$ exchange the value would be
$\approx$ 0.2.
Due to charge conjugation the $\eta^\prime$ can not
be produced by $\omega$-Pomeron exchange and Isospin
forbids
$\rho$-Pomeron exchange.
Therefore, it would appear that DPE is dominant in $\eta^\prime$
production.
For the $\eta$ there is no possibility of determining the
ratio because in the $\pi^+\pi^-\pi^0$ channel there is no
suitable reference signal.
\par
The cross section as a function of energy for the $J^{PC}$~=~$1^{++}$
$f_1(1285)$ and $f_1(1420)$ has been found to be constant~\cite{f1pap}.
Hence
both the $f_1(1285)$ and $f_1(1420)$ are consistent with being produced
by DPE~\cite{f1pap}.
For the  $J^{PC}$~=~$1^{++}$ states the model of
Close and Schuler predicts that
$J_Z$~=~$\pm1$ should dominate, which has been found experimentally to be
correct~\cite{f1pap}, and in addition that
\begin{equation}
\frac{d^3\sigma}{d\phi dt_1dt_2}\propto (\sqrt{t_2} - \sqrt{t_1})^2 +
4\sqrt{t_1 t_2}\sin^2 \phi/2
\label{1pp}
\end{equation}
Fig.~\ref{fi:1}c) and d) shows the
the $\phi$ distributions for the $f_1(1285)$ and
$f_1(1420)$. The distributions have been fitted to the form
$\alpha + \beta \sin^2 \phi/2 $, which describes the data well.
Equation~(\ref{1pp})
also predicts that when $|t_2 - t_1|$ is small
$d\sigma/d\phi$ should be proportional to $\sin^2 \phi/2 $ while
when
$|t_2 - t_1|$ is large
$d\sigma/d\phi$ should be constant. Fig.~\ref{fi:1}e) and f)
show the $\phi$ distributions for the $f_1(1285)$ for
$| t_1 - t_2 |$ $\le$ 0.2~GeV$^2$
and $|t_1 - t_2|$ $\ge$ 0.4~GeV$^2$ respectively;
as can be seen from the figures the expected trend is
observed in the data.
\par
The $f_0(980)$, $f_0(1500)$ and $f_2(1270)$
are other states for which the
cross section as a function of energy
has been found to be constant~\cite{pipikk} and hence are consistent
with being produced by DPE.
For the scalar states and for the tensor states with $J_Z$~=~0
Close and Schuler have predicted that
\begin{equation}
\frac{d\sigma}{d\phi }\propto (R  - \cos \phi)^2
\label{2pp}
\end{equation}
where
R is predicted from ref.~\cite{clschul}
to be a function of  $t_1t_2$ and can be
negative or positive.
In order to compare the data with the model we have studied the
$\phi$ dependences for the $f_0(980)$, $f_0(1500)$ and
$f_2(1270)$ by studying their decays to \pipi.
\par
In ref.~\cite{pipikk} a Partial Wave Analysis (PWA)
was performed in six bins of $\phi$ in order to determine
the $\phi$ dependences of the above resonances.
It has not been possible to perform a PWA as
a function of $\phi$ and $t_1t_2$. Therefore
in order to determine the $\phi$ dependences
in intervals of $t_1t_2$ we have performed a
fit to the total mass spectrum in each interval
using the method described in ref.~\cite{oldpipi}.
A PWA has been performed in each $t_1t_2$ interval discussed below,
integrated over $\phi$, to determine the
amount of $f_2(1270)$ produced with $J_Z$~=~0 compared to
$J_Z$~=~$\pm1$. The amount of $J_Z$~=~$\pm1$ is found to be
$\approx$~10~\% of the
$J_Z$~=~$0$ contribution
irrespective of the $t_1t_2$ interval.
The $J_Z$~=~$\pm2$ contribution is consistent with zero.
\par
Fig.~\ref{fi:2}a), d) and g) show the $\phi$ distributions,
for the
$f_0(980)$, $f_0(1500)$ and $f_2(1270)$
respectively, for all the data.
These distributions are similar to those
found from a fit to the PWA amplitudes~\cite{pipikk}.
The $\phi$ distributions have been fitted to the
form given in equation~(\ref{2pp}). The values
of R determined from the fit are given in table~\ref{ta:a}.
Since R is predicted
to be a function of $t_1t_2$
the $\phi$ distributions have been analysed in
two different intervals of $t_1t_2$, the corresponding values of
R are given in table~\ref{ta:a}.
Fig.~\ref{fi:2}b), e) and h) show the $\phi$ distributions,
for the
$f_0(980)$, $f_0(1500)$ and $f_2(1270)$ respectively for
$|t_1t_2|$ ~$\leq$~0.01~GeV$^4$.
Fig.~\ref{fi:2}c), f) and i) show the corresponding
$\phi$ distributions
for
$|t_1t_2|$ ~$\geq$~0.08~GeV$^4$.
\par
For the resonances
studied to date, in the WA102 experiment,
the model of Close and Schuler~\cite{clschul} is in qualitative agreement
with the data.
{\em
Hence the data are consistent with the hypothesis that the
Pomeron transforms as a non-conserved vector current.
}
\par
In order to understand what happens if a different
particle is exchanged a study
has been made of
the reactions
\begin{equation}
pp \rightarrow \Delta^{++}_{f} (\pi^-) p_{s}
\label{eq:c}
\end{equation}
and
\begin{equation}
pp \rightarrow \Delta^{++}_{f} (\rho^-) p_{s}
\label{eq:cc}
\end{equation}
In this case a particle with $I$~=~1 has
to be exchanged from the $p$-$\Delta^{++}$ vertex and hence we are no longer
studying reactions which are DPE.
In the case of central $\pi^-$ production the most likely production
mechanism is $\pi$-Pomeron exchange. For the
$\rho^-$ production the most likely production mechanisms
are $\rho$-Pomeron and $\pi$-$\pi$ exchange.
\par
To select reaction~(\ref{eq:c}) a
study
has been made of the reaction
\begin{equation}
pp \rightarrow p_{f} (\pi^+ \pi^-) p_{s}
\label{eq:b}
\end{equation}
at 450~GeV/c.
The isolation of
reaction~(\ref{eq:b}) has been described in ref.~\cite{pipipap}.
Fig~\ref{fi:3}a) shows the
$p_f\pi^+$
effective mass spectrum  where a clear peak corresponding to the
$\Delta^{++}(1232)$ can be observed.
In order to separate reaction~(\ref{eq:c})
from the reaction
\begin{equation}
pp \rightarrow N^*_f p_{s}
\end{equation}
where $N^*_f \rightarrow \Delta^{++}_f \pi^-$,
the rapidity gap between the $\pi^-$ and the $p_f\pi^+$ system
has been required to be greater than 2.0 units.
The resulting
$p_f\pi^+$
effective mass spectrum  is shown in fig.~\ref{fi:3}b).
Reaction~(\ref{eq:c}) has been selected by requiring
$M(p_f\pi^+)$~$\leq$~1.4~GeV.
The remaining $p_f\pi^+\pi^-$ and $\pi^+\pi^-$ mass spectra
have no resonance contributions.
\par
The four momentum transfer ($|t_{fast}|$) at the beam-$\Delta^{++}$
vertex is shown in fig.~\ref{fi:3}c) and has been fitted to the
form
\begin{equation}
\frac{d\sigma}{dt} = \frac{\alpha |t|}{(|t| +
m_\pi^2)^2}e^{-2\beta(|t|+m_\pi^2)}
\end{equation}
which is the standard expression used to describe $\pi$ exchange~\cite{piex}.
The first bin in the distributions has been excluded from the fit
due to the fact that the uncertainties in the acceptance correction
are greatest in this bin.
The fit describes the data well, and yields $\beta$~=~2.4~$\pm$~0.2~GeV$^{-2}$,
consistent with $\pi$ exchange~\cite{piex}.
\par
The four momentum transfer ($|t_{slow}|$) at the target-slow
vertex is shown in fig.~\ref{fi:3}d) and has been fitted to the
form $e^{-bt}$.
The data with $|t|$~$\leq$~0.1~GeV$^2$
has been excluded from the fit
due to the poor acceptance for the slow proton in this range.
The fit yields a value of $b$~=~6.2~$\pm$~0.1~GeV$^{-2}$
which is compatible with Pomeron exchange being the dominant
contribution~\cite{pomex}.
\par
In this case
the azimuthal angle $\phi$ is defined as the angle between the $p_T$
vectors of the slow proton and the $\Delta^{++}$ and is shown
in fig.~\ref{fi:3}e). The $\phi$ distribution is consistent with being
flat as would be expected
if the process was dominated by the exchange of a particle with spin 0,
as in $\pi$ exchange.
\par
To select reaction~(\ref{eq:cc}) a
study has then been made of the reaction
\begin{equation}
pp \rightarrow p_{f} (\pi^+ \pi^-\pi^0) p_{s}
\end{equation}
the isolation of which
has been described in ref.~\cite{pipipi0}.
Fig~\ref{fi:4}a) shows the
$p_f\pi^+$
effective mass spectrum  where a clear peak corresponding to the
$\Delta^{++}(1232)$ can be observed which
has been selected by requiring
$M(p_f\pi^+)$~$\leq$~1.4~GeV
in order to select the reaction
\begin{equation}
pp \rightarrow \Delta^{++}_{f} (\pi^-\pi^0) p_{s}
\end{equation}
A rapidity gap of 2.0 units is required between the
$\Delta^{++}$ and the $\pi^-\pi^0$ system.
\par
Fig.~\ref{fi:4}b) shows the
$\pi^-\pi^0$
effective mass spectrum  where a clear peak corresponding to the
$\rho^-(770)$ can be observed.
The mass spectrum
has been fitted
using two Breit-Wigners,
representing
the $\rho^-(770)$ and the broad enhancement in the 1.65~GeV region,
plus a background of the form
$a(m-m_{th})^{b}exp(-cm-dm^{2})$, where
$m$ is the
$\pi^{-}\pi^{0}$
mass,
$m_{th}$ is the
$\pi^{-}\pi^{0}$
threshold mass and
a, b, c, d are fit
parameters.
The fit yields for the $\rho^-(770)$ M~=~771~$\pm$~3~MeV,
$\Gamma$~=~160~$\pm$~15~MeV and for the 1.65~GeV region
M~=~1660~$\pm$~9~MeV,
$\Gamma$~=~240~$\pm$~25~MeV, which could be due to the $\rho_3(1690)$
or the $\rho(1700)$.
\par
In order to determine the
four momentum transfer ($|t|$) at the beam-$\Delta^{++}$
vertex for $\rho^-$ production
the $\pi^-\pi^0$ mass spectrum has been fitted in 0.05 GeV$^2$ bins
of $t$
with the parameters of the resonances fixed to those obtained from the
fits to the total data.
The resulting distribution
is shown in fig.~\ref{fi:4}c). In this case
it can not be fitted with the $\pi$ exchange formula and instead
has been fitted to the form $e^{-bt}$.
The first bin in the distributions has been excluded from the fit
due to the fact that the uncertainties in the acceptance correction
are greatest in this bin.
The fit yields a value of $b$~=~4.7~$\pm$~0.1~GeV$^{-2}$
which is compatible with $\rho$ exchange being the dominant
contribution~\cite{rhoex}.
\par
The four momentum transfer ($|t_{slow}|$) at the target-slow
vertex is shown in fig.~\ref{fi:3}d) and has been fitted to the
form $e^{-bt}$.
The data for $|t|$~$\leq$~0.1~GeV$^2$
has been excluded from the fit
due to the poor acceptance for the slow proton in this range.
The fit yields a value of $b$~=~6.1~$\pm$~0.1~GeV$^{-2}$
which is compatible with Pomeron exchange being the dominant
contribution~\cite{pomex}.
\par
In order to determine the
azimuthal angle $\phi$ between the $p_T$
vectors of the slow proton and the $\Delta^{++}$
for $\rho^-$ production
the $\pi^-\pi^0$ mass spectrum has been fitted in
30 degree bins of $\phi$
with the parameters of the resonances fixed to those obtained from the
fits to the total data.
The resulting distribution
is shown
in fig.~\ref{fi:4}e). The $\phi$ distribution is clearly not flat in
this case.
Hence the $\rho^-$ is consistent with being produced by particles
that carry spin, for example $\rho$-Pomeron exchange with the
Pomeron transforming like a non-conserved vector current.
\par
In summary,
for the resonances
studied to date which are compatible with
being produced by DPE,
the model of Close and Schuler~\cite{clschul} is in qualitative agreement
with the data and hence is consistent with the Pomeron
transforming like a non-conserved vector current.
When one of the particles exchanged is known to have spin 0, namely
$\pi$-Pomeron exchange, the $\phi$ distribution is flat.
When $\rho$-Pomeron
exchange is the dominant contribution the $\phi$ distribution
is not flat.
\begin{center}
{\bf Acknowledgements}
\end{center}
\par
This work is supported, in part, by grants from
the British Particle Physics and Astronomy Research Council,
the British Royal Society,
the Ministry of Education, Science, Sports and Culture of Japan
(grants no. 04044159 and 07044098), the French Programme International
de Cooperation Scientifique (grant no. 576)
and
the Russian Foundation for Basic Research
(grants 96-15-96633 and 98-02-22032).

\newpage
\newpage
\begin{table}[h]
\caption{Determination of R from
fits to the $\phi$ distributions.}
\label{ta:a}
\vspace{1in}
\begin{center}
\begin{tabular}{|c|c|c|c|} \hline
 & & & \\
 &All & $|t_1t_2|$ $\leq$ 0.01 GeV$^4$ &$|t_1t_2|$ $\geq$ 0.08 GeV$^4$\\
 &data & &\\
 & & & \\ \hline
 & & & \\
$f_0(980)$  &-1.9 $\pm$ 0.1 & -2.3 $\pm$ 0.3   &0.01 $\pm$ 0.03 \\
 & & & \\ \hline
 & & &  \\
$f_0(1500)$  &-1.6 $\pm$ 0.2  & -1.7 $\pm$ 0.2  &0.16 $\pm$ 0.06  \\
 & & & \\ \hline
 & & & \\
$f_2(1270)$  &2.5 $\pm$ 0.1  & 2.1 $\pm$ 0.2 &3.8 $\pm$ 0.4\\
 & & & \\ \hline
\end{tabular}
\end{center}
\end{table}
\clearpage
{ \large \bf Figures \rm}
\begin{figure}[h]
\caption{The azimuthal angle $\phi$ between the fast and
slow protons for
a) the $\eta$, b) the $\eta^\prime$, c) the $f_1(1285)$ and
d) the $f_1(1420)$.
The $\phi$ distribution for the $f_1(1285)$ for
e) $| t_1 - t_2 |$ $\le$ 0.2 GeV$^2$
and f) $|t_1 - t_2|$ $\ge$ 0.4 GeV$^2$.
The fits to the distributions are described in the text.
}
\label{fi:1}
\end{figure}
\begin{figure}[h]
\caption{The azimuthal angle $\phi$ between the fast and
slow protons.
The first column is for all the events, the second for
events with  $|t_1t_2|$~$\leq$~0.01~GeV$^{4}$ and
the last for those with
$|t_1t_2|$~$\geq$~0.08~GeV$^{4}$ for the
$f_0(970)$ (a, b and c), the
$f_0(1500)$ (d, e and f) and for the
$f_2(1270)$ (g, h and i) respectively.
The fits to the distributions are described in the text.
}
\label{fi:2}
\end{figure}
\begin{figure}[h]
\caption{The selection of the reaction
$pp \rightarrow \Delta^{++}_{f} (\pi^-) p_{s}$.
The $p_f\pi^+$ effective mass spectrum for a) all the data
and b) for events with a rapidity gap greater than 2.0 units
between the $\pi^-$ and $p_f\pi^+$ system.
c) The four momentum transfer distribution at the beam-$\Delta^{++}$
vertex with a fit to $\pi$ exchange described in the text.
d) The four momentum transfer distribution at the target-slow
vertex with a fit described in the text.
e) The azimuthal
angle $\phi$ between the $p_T$
vectors of the slow proton and the $\Delta^{++}$.
}
\label{fi:3}
\end{figure}
\begin{figure}[h]
\caption{The selection of the reaction
$pp \rightarrow \Delta^{++}_{f} (\pi^-\pi^0) p_{s}$.
a) The $p_f\pi^+$ effective mass spectrum.
b) The $\pi^-\pi^0$ effective mass spectrum with fit described in the text.
c) The four momentum transfer distribution at the beam-$\Delta^{++}$
vertex for $\rho^-$ production with a fit described in the text.
d) The four momentum transfer distribution at the target-slow
vertex for $\rho^-$ production with a fit described in the text.
e) The azimuthal
angle $\phi$ between the $p_T$
vectors of the slow proton and the $\Delta^{++}$ for $\rho^-$ production.
}
\label{fi:4}
\end{figure}
\newpage
\begin{center}
\epsfig{figure=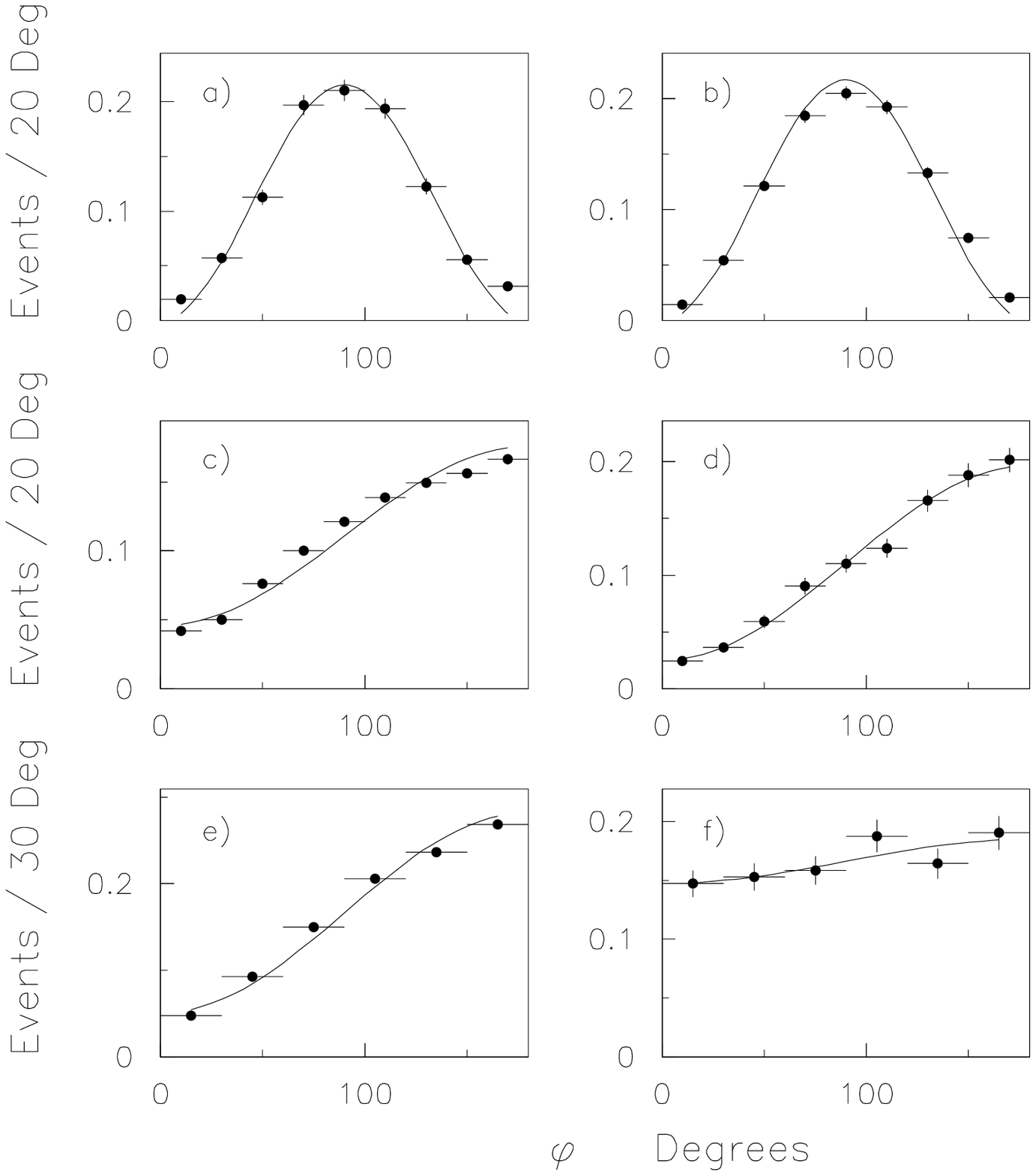,height=22cm,width=17cm}
\end{center}
\begin{center} {Figure 1} \end{center}
\newpage
\begin{center}
\epsfig{figure=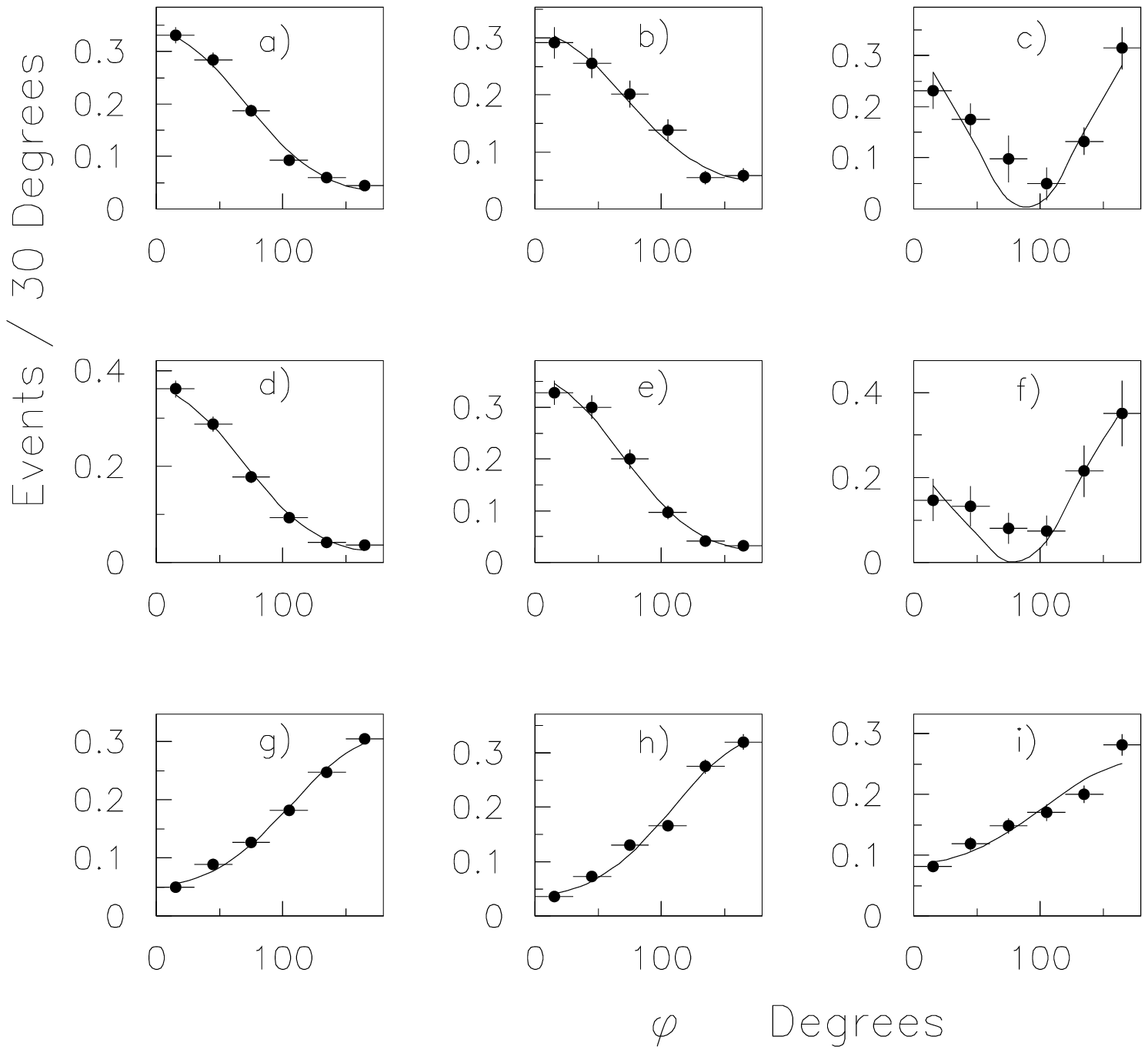,height=22cm,width=17cm}
\end{center}
\begin{center} {Figure 2} \end{center}
\newpage
\begin{center}
\epsfig{figure=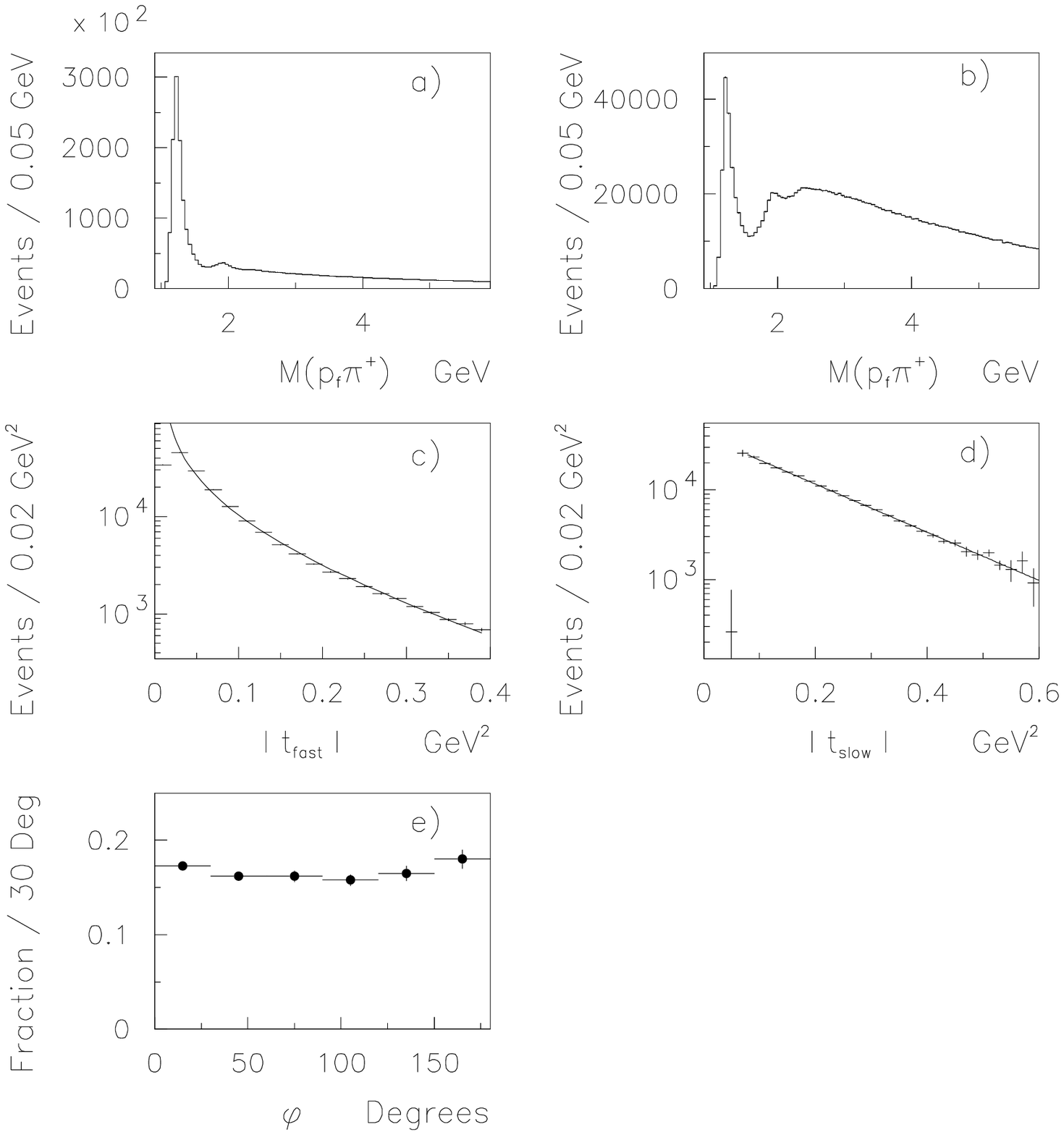,height=22cm,width=17cm}
\end{center}
\begin{center} {Figure 3} \end{center}
\newpage
\begin{center}
\epsfig{figure=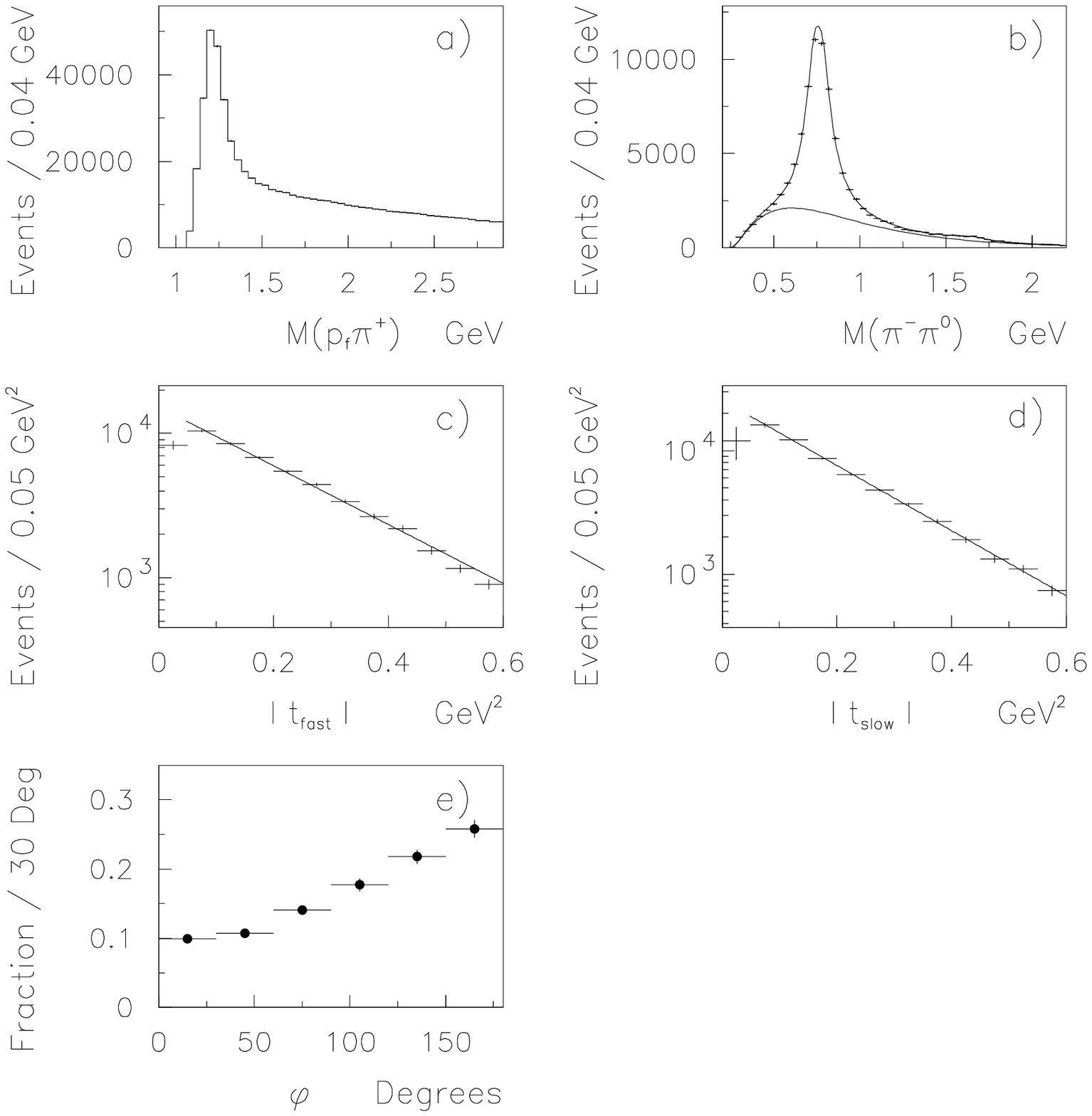,height=22cm,width=17cm}
\end{center}
\begin{center} {Figure 4} \end{center}

\bigskip
\newpage
\begin{thebibliography}{99}
\bibitem{WADPT}
D. Barberis {\em et al.,} Phys. Lett. {\bf B397 } \rm (1997) 339.
\bibitem{closeak}
F.E. Close and A. Kirk, Phys. Lett. {\bf B397 } \rm (1997) 333.
\bibitem{memoriam}
A. Kirk, Yad. Fiz. {\bf 62} (1999) 439.
\bibitem{ourpubs}
D. Barberis {\em et al.,} Phys. Lett. {\bf B432} (1998) 436;\\
D. Barberis {\em et al.,} Phys. Lett. {\bf B436} (1998) 204.
\bibitem{0mppap}
D. Barberis {\em et al.,} Phys. Lett. {\bf B427} (1998) 398.
\bibitem{f1pap}
D. Barberis {\em et al.,} Phys. Lett. {\bf B440} (1998) 225.
\bibitem{pipikk}
D. Barberis {\em et al.,} hep-ex/9907055. To be published in Phys. Lett.
\bibitem{pipipi0}
D. Barberis {\em et al.,} Phys. Lett. {\bf B422} (1998) 399.
\bibitem{angdist}
F.E. Close Phys. Lett. {\bf B419}  (1998) 387; \\
P. Castoldi, R. Escribano and J.-M. Frere,
Phys. Lett. {\bf B425} (1998) 359; \\
N. I. Kochelev, hep-ph/9902203; \\
F.E. Close and G. Schuler, Phys. Lett. {\bf B458} (1999) 127; \\
N. I. Kochelev, T. Morii and A.V. Vinnikov, hep-ph/9903279.
\bibitem{clschul}
F.E. Close and G. Schuler hep-ph/9905305.
\bibitem{eiota}
F.E. Close and A. Kirk, Zeit. Phys. {\bf C76 } \rm (1997) 469.
\bibitem{WA76etapipi}
T.A. Armstrong {\em et al.,} Proceedings of the EPS-HEP 85 (1985) 314.
\bibitem{oldpipi}
T.A. Armstrong {\em et al.,} Zeit. Phys. {\bf C51} (1991) 351.
\bibitem{pipipap}
D. Barberis {\em et al.,} Phys. Lett. {\bf B453} (1999) 316.
\bibitem{piex}
D. Alde {\em et al.,} Zeit. Phys. {\bf C54 } (1992) 553.
\bibitem{pomex}
S.N. Ganguli and D.P. Roy, Phys. Rep. {\bf 67} (1980) 203.
\bibitem{rhoex}
M.L. Perl {\em High Energy Hadron Physics} Wiley Press (1974).
\end{thebibliography}
\end{document}